# Use of Solr and Xapian in the Invenio document repository software

Patrick Glauner, Jan Iwaszkiewicz , Jean-Yves Le Meur and Tibor Simko, for Invenio collaboration

```
European Organization for Nuclear Research
            CERN, IT Department
       CH-1211, Geneva 23, Switzerland
    {patrick.oliver.glauner, jan.iwaszkiewicz,
     jean-yves.le.meur, tibor.simko}@cern.ch
               http://www.cern.ch
```

## Abstract

Invenio is a free comprehensive web-based document repository and digital library software suite originally developed at CERN. It can serve a variety of use cases from an institutional repository or digital library to a web journal. In order to fully use full-text documents for efficient search and ranking, Solr was integrated into Invenio through a generic bridge. Solr indexes extracted full-texts and most relevant metadata. Consequently, Invenio takes advantage of Solr's efficient search and word similarity ranking capabilities. In this paper, we first give an overview of Invenio, its capabilities and features. We then present our open source Solr integration as well as scalability challenges that arose for an Invenio-based multi-million record repository: the CERN Document Server. We also compare our Solr adapter to an alternative Xapian adapter using the same generic bridge. Both integrations are distributed with the Invenio package and ready to be used by the institutions using or adopting Invenio.

## 1  Invenio

Invenio[1] is a free Pythonic software suite to run a web-based document repository or digital library. It has been originally developed at CERN, the European Organization for Nuclear Research[2], in 2002 to provide the CERN Document Server[3]. It is intended to run mid to large-size repositories containing typically 100K to 10M records which can be of any kind such as articles, books or multimedia content. It is nowadays co-developed by an international collaboration with contributions from institutions such as Cornell University, École Polytechnique Fédérale de Lausanne, GSI Helmholtz Centre for Heavy Ion Research, Harvard-Smithsonian Center for Astrophysics or SLAC National Accelerator Laboratory. Currently, there are above 30 mid to large-size Invenio-based repositories including INSPIRE[4] and ILO Labordoc[5].

Invenio is based on the Linux, Apache, MySQL, Python (LAMP) stack and uses MARC 21 as its underlying bibliographic format. Its modular design allows it to serve a broad range of use cases from an institutional or object repository to a digital library or web journal. It offers a variety of features such as a navigable multi-dimension collection tree, a powerful search engine, customizable metadata, automatic reference extraction, interaction with other services, user personalization and standard bibliographic output formats.

Data can be ingested in an Invenio repository through different channels. First, users can submit data through customizable web interfaces of the `WebSubmit` module. Submissions can be configured by administrators and implemented by developers. Second, data can be harvested from OAI-PMH compliant repositories. Third, using XSLT or Invenio's `BibConvert` language, records in different formats can be ingested. Further ingestion channels include submissions via emails and a lightweight API that can be used by robots. Also, users can claim authorship of existing records to add them to their collections. There are several tools to curate records in an Invenio instance such as multi-record modificator, duplicate record finder and record merger. Invenio keeps a revision history of records and provides rich web interfaces to make these features available to catalogers. Additionally, Invenio offers a variety of dissemination facilities including alerts, baskets, comments or export to different formats. Invenio's display layer is adaptable thanks to its templating approach. The appearance and formatting of an instance's collections and records can be customized and configured. This broadens the range of Invenio use cases. Invenio is offered in more than 20 languages permitting users to switch language on the fly [6].

Invenio's indexing module `BibIndex` creates indexes of metadata fields which are subsequently stored in the database. They are designed to speed up queries assuming a high amount of selects and lower amount of updates. `WebSearch` is Invenio's search engine module using the previously created indexes for efficient search. It supports different types of user queries including words, phrases, regular expressions or second order "cited by" operators. The Invenio ranking module `BibRank` offers multiple ranking methods including word similarity ranking and citation ranking. Both search and ranking are two separate steps. The searcher returns the search results which are subsequently used by the ranker. This permits to execute very efficient search queries and to use independent technology for ranking [7].

## 2  The CERN Document Server

The CERN Document Server (CDS) is one of the largest Invenio instances managing more than 1.5M records and 400K full-text documents in high energy physics domain. It receives more than 10K unique visits per day with occasional peaks of up to 350K visits per day during important physics announcements. It serves as reference repository for the development of the features described in this paper. The CDS environment consists of a load balancer, six worker nodes and two database nodes, aside from storage nodes and a distributed full-text storage system to handle occasional high peaks of usage.

## 3  Solr and Xapian integrations

Invenio came with a native full-text indexer that was optimised for fast word queries but lacked advanced word proximity or semantic features. The goal of this work was to take advantage of existing external information retrieval systems for efficient and scalable full-text search and word similarity ranking. Different information retrieval systems such as Solr[8], elasticsearch[9] and Xapian[10] haven been evaluated. Solr has been most promising to be integrated for both its performance and several features such as facets, snippets, word similarity ranking and semantic search extensions. It was integrated in Invenio through a generic bridge in the respective modules `BibIndex`, `WebSearch` and `BibRank`. The generic bridge allows Invenio installations to easily plug any suitable third-party search and ranking engine by implementing an adapter – see Figure 1.

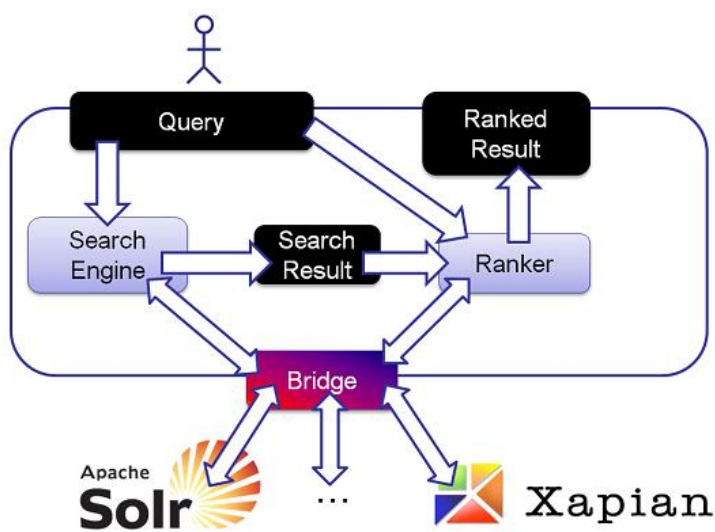

Figure 1: External search and word similarity ranking through the bridge

First, Solr indexes the extracted full-texts of the Invenio instance's records for the use case of full-text search. Second, most important metadata such as titles, abstracts or keywords are re-used for the use cases of external word similarity ranking and finding records similar to a specific record. `WebSearch` internally stores search result record identifiers in an `intbitset`[11], a highly-optimized Invenio data structure. Tests performed with different technologies have shown that passing the entire result set – potentially containing hundreds of thousands record identifiers – from Solr to Invenio would take over a second which was not acceptable. This led to the development of a Solr `intbitset` extension[12] in collaboration with the Astrophysics Data System team (ADS)[13]. The extension permits Solr to return the record identifiers of the entire result set in the form representing serialized binary Python `intbitset` structure and allows transmission of 1M identifiers from Solr to Invenio in about 0.1 seconds.

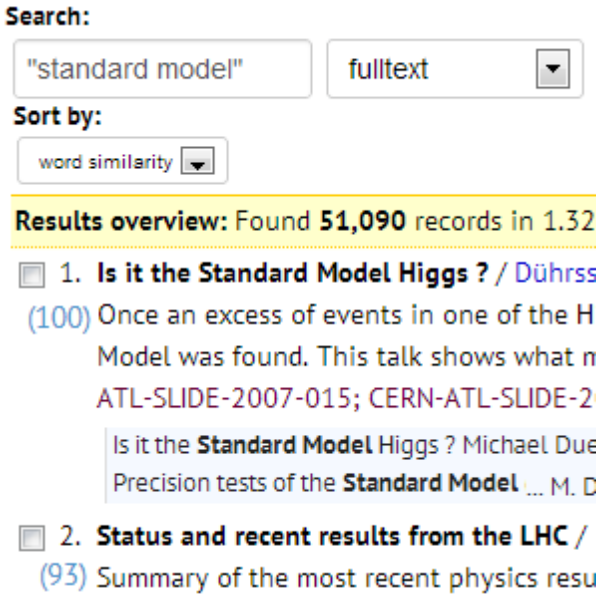

Figure 2: Solr-based word similarity result by percentage

The word similarity ranking part of the Solr adapter – see Figure 2 – ranks the indexed metadata and full-text depending on weights set by Invenio instance administrators. For typical Invenio ranking use cases, only the top results need to be retrieved. As Invenio ranking relies on previously retrieved search result, the Solr ranker needs to retrieve the search result hitset initially. Passing the potentially very large hitset via HTTP could then slow the communication down. For queries resulting in more than 10K hits, only the identifiers of the 10K latest added records are passed. This might not return an optimal solution, but a close approximation of user expectations since latest publications are usually more relevant in most Invenio use cases. This parameter can be configured in each Invenio instance.

Finding similar records to a specific record is supported via the Solr search handler MoreLikeThis (MLT)[14]. During tests on the CDS corpus, full-text was observed to distort MLT results. Eventually, the MLT index containing abstract, author, keywords and title was found to return the best results on CDS. In order to increase performance, Solr only evaluates the first 1K tokens. This and further MLT parameters can be configured by administrators in Invenio.

The CERN Document Server's Solr instance runs on a dedicated server node. Its index size is about 40 GB. Solr performance – see Table 1 – was found good for the CDS corpus with linear tendency for indexing time, index size and querying time [15]. Also, multi-user scalability was found good under both average and high user request load conditions.

|  | Solr | Xapian |
|---|---|---|
| Search result count | 16,903 | 15,945 |
| Search [sec] | 0.06 | 0.56 |
| Ranked top 10 [sec] | 0.06 | 0.40 |
| Ranked top 100 [sec] | 0.12 | 0.42 |
| Ranked top 1K [sec] | 0.22 | 0.45 |
| Ranked 10K [sec] | 1.46 | 0.70 |

Table 1: "higgs boson" full-text phrase search and ranking for the CDS corpus

A second adapter integrating Xapian was implemented. Since Xapian creates a separate database per index, the ranking part of the adapter has to query each index individually and subsequently aggregate the results. On the one hand, Solr returns top results faster, but goes exponentially slower for larger ranking result sets.

On the other hand, Xapian is slower in returning top results but offers more balanced performance since result speed decreases only slightly for larger ranking result sets.

## 4 Conclusions

Invenio is a comprehensive and powerful document repository and digital library software suite. It is intended to serve mid to large-size repositories containing a variety of records types. In order to fully use full-text for efficient search and ranking, Solr was integrated with an adapter through a generic search and word similarity ranking bridge. The performance of the Solr adapter was found good for the CERN Document Server corpus with linear tendency for indexing time, index size and query time, with satisfying multi-user scalability. Xapian has been evaluated and offers different strengths with a notably better response time in case large ranking result sets have to be retrieved at once. Both Solr and Xapian integrations are distributed with the Invenio package and ready to be used by the institutions and projects using or adopting Invenio.

## Acknowledgements

The present work benefited from the input of the Invenio collaboration. Special thanks go to Jerome Caffaro and Samuele Kaplun for many helpful discussions. The authors wish to thank the ADS team for its contribution to the Solr `intbitset` extension.